\documentclass[runningheads,a4paper]{llncs}

\usepackage{amssymb}
\usepackage{url}
\usepackage{color,epsf,epsfig,graphicx,times,url,float,listings,multirow}
\usepackage{epstopdf}
\usepackage{subfig}
\usepackage{tikz,ifthen,xkeyval,calc}

\usetikzlibrary{shapes,snakes}

\urldef{\mails}\path|{garlan, vdwivedi, iruchkin, schmerl}@cs.cmu.edu|
\newcommand{\keywords}[1]{\par\addvspace\baselineskip
\noindent\keywordname\enspace\ignorespaces#1}

\begin{document}

\mainmatter  

\title{Foundations and Tools for End-User Architecting}

\titlerunning{End-User Architecting}

%
%
\author{David Garlan\and Vishal Dwivedi\and Ivan Ruchkin\and Bradley Schmerl}
\authorrunning{End-User Architecting}

\institute{School of Computer Science \\
Carnegie Mellon University \\
 Pittsburgh, USA.\\
\mails
}

\toctitle{End-User Architecture}
\tocauthor{Garlan et al.}
\maketitle

\begin{abstract}
Within an increasing number of domains an important emerging need is the ability for technically  na\"{\i}ve users to compose computational elements into novel configurations. Examples include astronomers who create new analysis pipelines to process telescopic data, intelligence analysts who must process diverse sources of unstructured text to discover socio-technical trends, and medical researchers who have to process brain image data in new ways to understand disease pathways. Creating such compositions today typically requires low-level technical expertise, limiting the use of computational methods and increasing the cost of using them. In this paper we describe an approach --- which we term \emph{end-user architecting} --- that exploits the similarity between such compositional activities and those of software architects. Drawing on the rich heritage of software architecture languages, methods, and tools, we show how those techniques can be adapted to support end users in composing rich computational systems through domain-specific compositional paradigms and component repositories, without requiring that they have knowledge of the low-level implementation details of the components or the compositional infrastructure.  Further, we outline a set of open research challenges that the area of end-user architecting raises.

\keywords{end-user architecture, end-user architecting, software architecture, end-user programming, software composition, software development tools.}
\end{abstract}

\section{Introduction}
\label{sec:intro}

Increasingly users rely on computation to support their professional activities. In some cases turnkey applications and services are sufficient to carry out computational tasks. However, in many situations users must adapt computing to their specific needs. These adaptations can take many forms: from setting preferences in applications, to ``programming''  spreadsheets, to creating orchestrations of services in support of some business process.  This situation has given rise to an interest in end-user programming~\cite{Nardi160165}, and, more generally, end-user software engineering~\cite{KoABBBESLLMRRSW11} or end-user computing~\cite{goodell_end-user_1997}. This emerging field attempts to find ways to better support users who, unlike professional programmers, do not have deep technical knowledge, but must somehow find ways to harness the power of computation to support their tasks.

One important subclass of end-user computation arises in domains where users must compose existing computational elements into novel configurations. Examples include e-science (e.g., astronomers who create new analysis pipelines to process telescopic data), intelligence analysis (e.g., policy planners who process diverse sources of unstructured text to discover socio-technical trends), and medicine (e.g., researchers who process repositories of brain imaging data to discover new disease pathways).

\begin{figure}
\vspace{-20pt}
\centering
\begin{tabular}{cc}
\epsfig{file=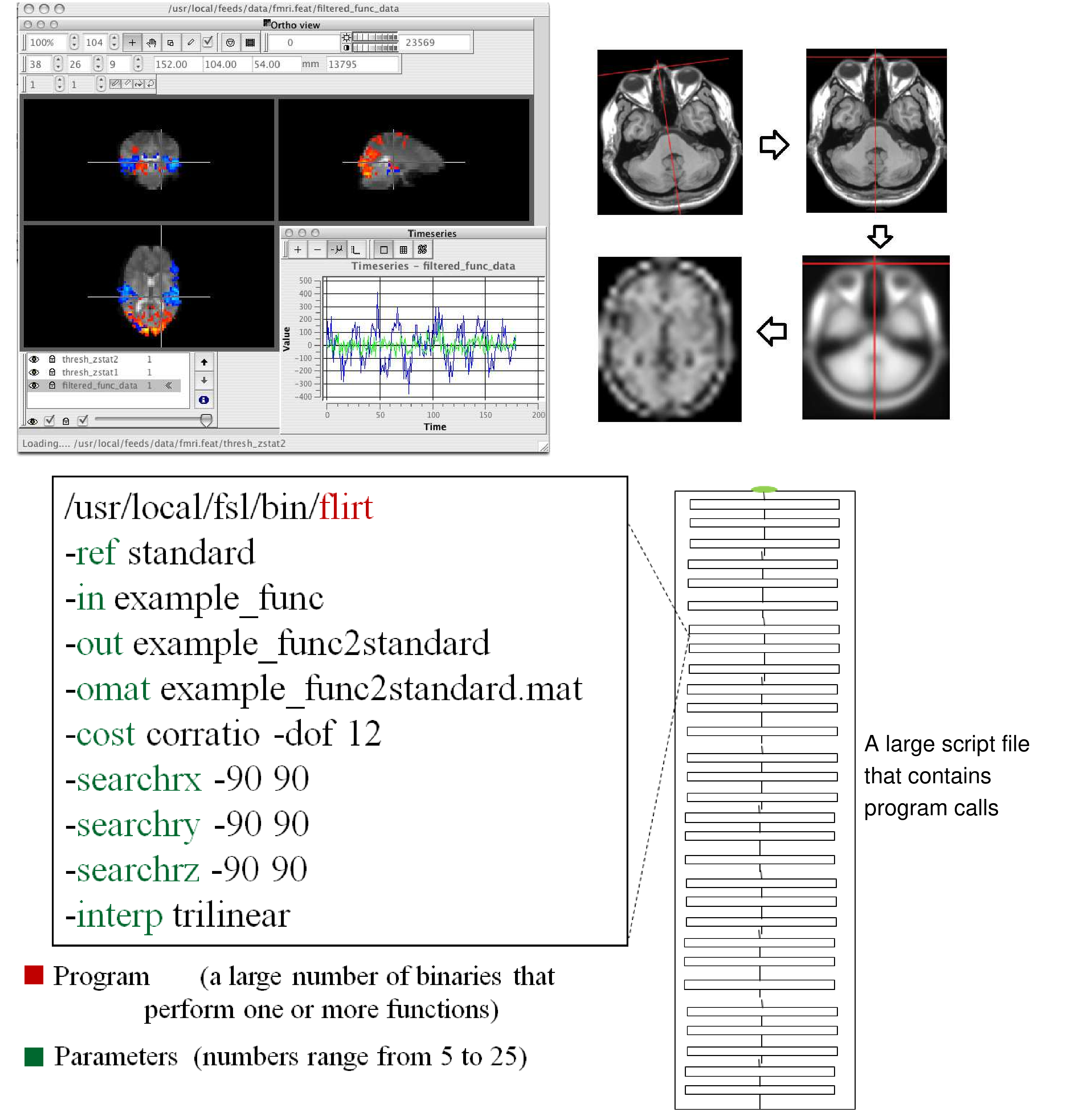,width=0.55\linewidth}
\end{tabular}
\vspace{-10pt}
\caption{Compositions in the neuroscience domain.}
\vspace{-20pt}
\label{fig:neuroscience}
\end{figure}

In these domains professionals typically have access to a large number of existing applications and data sets, which must be composed in novel ways to gain insight, carry out ``what if'' experiments, generate reports and research findings, etc. For example, in the field of brain imaging, scientists study samples of brain images and neural activity to diagnose disease patterns. Innovative research in this domain requires that scientists compose a large number of tools and apply them to brain-imaging data sets to diagnose problems, such as malformations and structural or functional deformities. There also exist dozens (if not hundreds) of brain image processing tools for image recognition, image alignment, filtering, volumetric analytics, mapping, etc. Figure~\ref{fig:neuroscience} illustrates a popular neuroscience tool suite, called FSL, that is used to create scripts for analyzing FMRI~\cite{pekar2006} data.
%

Unfortunately, assembling such elements into coherent compositions is a non-trivial matter. In many cases users must have detailed low-level knowledge of things like application parameter settings, application invocation idiosyncrasies, file locations and naming conventions, data formats and encodings, ordering restrictions, and scripting languages. In Figure~\ref{fig:neuroscience}, for example, users must create and execute detailed scripts illustrated at the bottom of the figure.

Further, it may be difficult for end users to determine whether a set of components can be composed at all, and, if not, what to do about it. For example, differences in data encodings may make direct component composition infeasible without the inclusion of one or more format converters. Even when a legal composition can be achieved, it may not have the performance (or other quality attributes) critical to the needs of the end user. And, even when a suitably performing composition can be created, it may be difficult to share it with peers or reuse it in similar but different settings.

In this paper we advocate an approach to these problems that exploits the similarity between such compositions and software architecture, and attempts to leverage the considerable advances made within that field over the past two decades. The key idea is to view the activities of these end users as engaging in architectural design within a domain-specific style and to represent those architectures explicitly. As we will see, such explicit representation allows one to raise the level of abstraction for composition, provide criteria for evaluating the soundness and quality of a composition, support reuse and parametrization, and establish a platform for a host of task-enhancing services such as program synthesis, analysis, compilation, execution, and debugging.

By approaching the problem in this way we identify a new field of concern, which we term \emph{end-user architecting}. Similar to end-user programming~\cite{Nardi160165}, it recognizes up front that the key issue is bridging the gap between available computational resources and the skill set of the users who must harness them --- users who typically have weak or non-existent programming skills.
But unlike end-user programming, it seeks to find higher-level abstractions that leverage the considerable advances in software architecture languages, methods, and tools to support component composition, analysis and execution.

In Section~\ref{sec:problem} we revisit the problem, highlighting the cross-cutting similarities in computing needs for composition-based domains such as those mentioned above, and we outline the challenges for solving the problems of users in these domains. Section~\ref{sec:endUserArchitecture} makes the case for taking an architectural perspective on the problem, and outlines an approach in which software architecture tools and techniques can be incorporated into environments that support end-user architecting. Section~\ref{sec:caseStudies} illustrates how this approach can be applied by considering three case studies. Section~\ref{sec:relatedWork} considers related work, and Section~\ref{sec:discussion} explores some of the open research challenges in this area.

\section{The Problem}\label{sec:problem}

As noted above, an increasing number of domains are evolving to depend on composing existing components to support their tasks. Table~\ref{tab:domains} lists examples of these domains, including  e-science, business processing, social science research, and electronic music synthesis.

\begin{table}[h!]
\centering
\begin{tabular}{|l|l|}
\hline
\bfseries Type & \bfseries Compositions\\
\hline
\hline
            \emph{Astronomy} & electromagnetic image processing tasks \cite{Deelman2005}	\\
\hline
            \emph{Bioinformatics} & biological data-analysis services  \cite{Letondal2005}\\
\hline
            \emph{Digital music production} & audio sequencing and editing { \cite{mcConahy2011}}  \\
\hline
            \emph{Environmental Science} & spatio-temporal experiments  	 \cite{VillaAR09} \\
\hline
            \emph{Geospatial Analysis} &  interactive visualization of geographical data \cite{MooreOzone11} \\
\hline
            \emph{Home Automation} & home devices and services { \cite{smartSpaces}}  \\
\hline
            \emph{Neuroscience} & brain-image processing libraries		 	 	 { \cite{Dwivedi11}} \\
\hline
            \emph{Scientific computing} & transformational workflows 		 { \cite{Segal07}} \\
\hline
            \emph{Socio-technical Analysis} &  dynamic network creation, analysis, reporting and simulation \cite{SchmerlGDBC11} \\
\hline
\end{tabular}
\vspace{6pt}
\caption{Domains involving end user compositions.}
\label{tab:domains}
\end{table}

While very different in their specific tasks and goals, the use of computation within these communities shares a number of common properties. First, it relies on compositions of existing components to accomplish computational tasks. For example, there exist large repositories of reusable components such as BioCatalogue~\cite{biocatalogue} for life science web services, the BIRN Data Repository~\cite{birn} for neuroscience data and analysis tools, and myExperiment~\cite{myExperiment} for scientific workflows.

Second, in many cases those compositions are complex, involving dozens of components, possibly running on many hosts. Thus, creating new compositions becomes a non-trivial task, often taking weeks to develop, test, and execute.

Third, quality attributes matter. While the specific quality attributes of concern vary from domain to domain, they typically include things like performance (time to complete a task), resource requirements (numbers of processors, storage requirements), availability (likelihood of crashing), privacy and security (protection of data). For example, a brain imaging composition may be of little use to a neuroscience researcher if it takes a week to execute, fails frequently, or compromises the privacy of the data.

Fourth, the socio-technical ecosystem within which these computations are used is complex, involving many roles and incentives~\cite{HowisonH11}. For example, \emph{researchers} care that their compositions produce credible outputs and that they can share their computations with their peers; \emph{component providers }care that they are given credit for the use of their components; \emph{regulators and funders} care that the provenance of all results is fully documented.

\begin{figure} [h!]
\begin{center}
\includegraphics[trim=11.2cm 0cm 0cm 0cm, clip=true, totalheight=0.25\textheight]{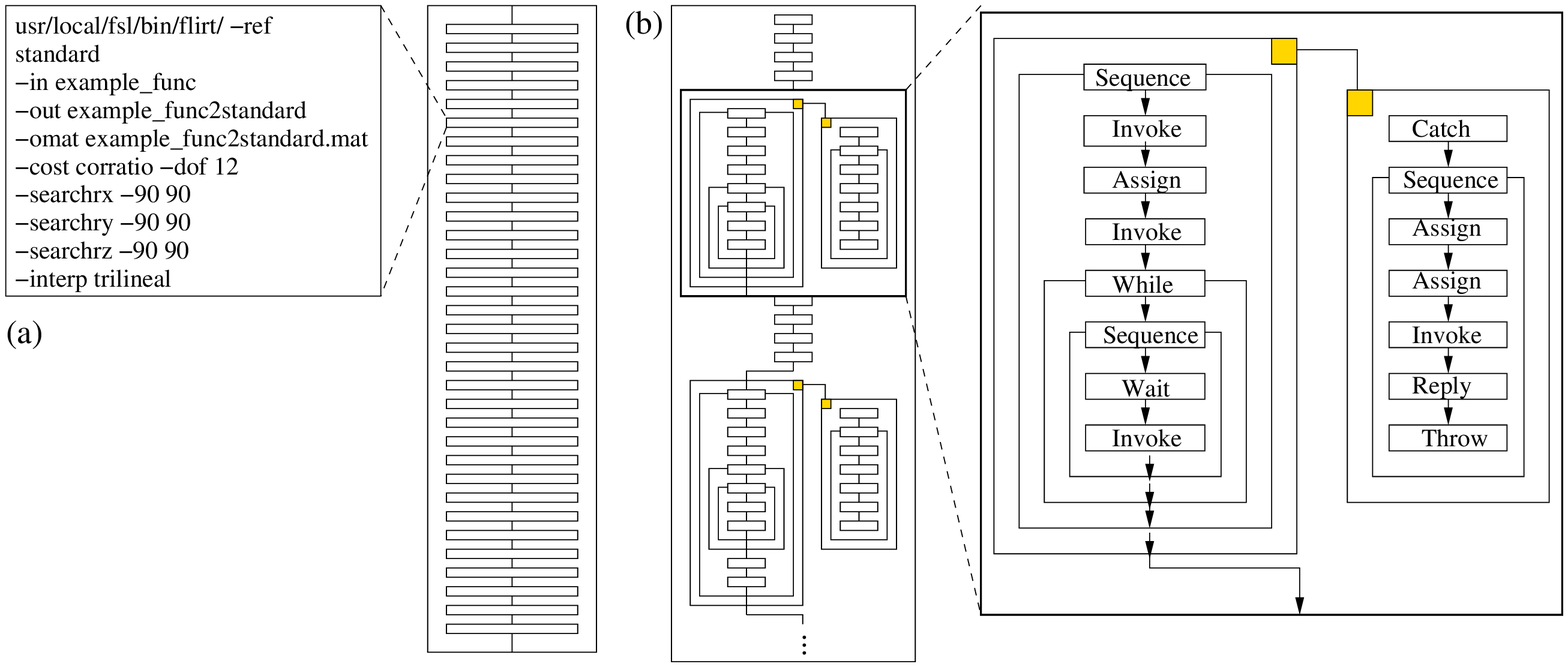}
\caption{A segment of BPEL orchestration of a socio-cultural analysis workflow.}
\label{fig:BPEL}
\end{center}
\end{figure}

Today these end-user communities are not well served by existing technology and development platforms. In particular, we can identify five critical barriers.
\begin{enumerate}
\item \textbf{Excessive technical detail:} Creating compositions today often requires knowledge of myriad low-level technical details, such as data formats, parameter settings, file locations, ordering constraints, execution conventions, scripting languages, etc. As Figure~\ref{fig:neuroscience} illustrates, brain imaging research using FSL tools requires a user to understand and create detailed execution scripts that specify how to configure each of the constituent tools, which may have dozens of configuration parameters. As another example, in the domain of intelligence analysis (cf. Section~\ref{sec:caseStudies}) a typical composition that involves two logical steps, but is executed in the context of a service-oriented architecture (SOA), requires the end user to specify a Business Processing Event Language (BPEL) script shown in Figure~\ref{fig:BPEL}~\cite{SchmerlGDBC11}. The script requires the user to explicitly specify low-level details that handle control flow, variable assignment, exception handling, and other programming constructs.
\item \textbf{Inappropriate computational models:} The computational models provided by typical execution platforms, such as SOA, may require end users to map their tasks into a computational vocabulary that is quite different from the natural way of decomposing the task in that domain. For example, tasks that are logically represented in the end user's mind as a workflow may have to be translated into the very-different vocabulary of service invocations executing on a SOA, as illustrated in Figure~\ref{fig:BPEL}.
\item \textbf{Inability to analyze compositions:} There may be many restrictions on legal ways to combine elements, dictated by things like format compatibility, domain-specific processing requirements, ordering constraints, and access rights to data and applications. Today, discovering whether a composition satisfies these restrictions is largely a matter of trial and error, since there are few tools to automate such checks. Moreover, even when a composition does satisfy the composition constraints, its extra-functional properties --- or quality attributes --- may be uncertain. For example, determining how long a given computation will take to produce results on a given data set can often be determined only by time-consuming experimentation.
\item \textbf{Lack of support for reuse:} An important requirement in many communities is the ability for professionals to share their compositions with others in those communities. For instance, brain researchers may want to replicate the analyses of others, or to adapt an existing analysis to a different setting (e.g., executed on different data sets). Packaging such compositions in a reusable and adaptable form is difficult, given the low-level nature of their encodings, and the brittleness of the specifications.
\item \textbf{Impoverished support for execution.} The execution environment for compositions is often impoverished. Compared to the capabilities of modern programming environments, end users have relatively few tools for things like compilation into efficient deployments, interactive testing and debugging (e.g., setting breakpoints, monitoring intermediate results, etc.), history tracking, and graceful handling of run-time errors. This follows in part from the fact that in many cases compositions are executed in a distributed environment using middleware that is not geared towards interactive use and exploration by technically naive users.
\end{enumerate}


This gap between the needs of end users and today's technology has a number of serious consequences. The cost of producing effective compositions is excessive because end users must become experts in implementation details not relevant to their primary task. The quality is low because compositions tend to be brittle and in many cases fail to meet their extra-functional requirements. Compositions are difficult to reuse, modify, and maintain, leading to gratuitous reinvention.

Recognizing these problems, a number of research- and practitioner-based efforts have produced platforms that provide end-user tools for composition, reuse and execution within specific domains. As described in more detail in Section~\ref{sec:relatedWork}, this is typically done through the creation of component repositories, and composition environments that support computational models appropriate to the domain, such as workflow execution, widget composition, data exploration or music synthesis and composition. Examples include Taverna for life sciences, the Ozone Widget Framework (OWF) for geospatial analysis, VisTrails for data exploration and visualization, Steinberg's Virtual Studio Technology (VST) for composing music effects, etc.

While many of these platforms have been quite successful, and several are in wide\-spread use, they are typically handcrafted for specific communities and domains --- often at great cost in development time and effort. What is needed, we would argue, is a foundational understanding of the problem and a general approach to a solution that gets at the heart of the mismatch between end user needs and technologies that must be exploited. Such foundations would ideally lead to a systematic approach to developing tools that surmount the barriers outlined earlier. In the next section we outline such an approach.

\section{End-User Architecture}\label{sec:endUserArchitecture}

The key to solving the problems outlined above is to recognize that the computational design activities performed by those communities are fundamentally architectural in nature. Recognizing that, one can then explore how modern techniques and tools in support of \emph{software architecture} can be applied to this new area of \emph{end-user architecting}.


Software architecture emerged as a subfield of software engineering in the 1990s as a way to tackle the increasing complexity of software systems design. While there are many definitions of software architecture, a typical one is~\cite{clements_documenting_2010}:
\begin{quotation}
\emph{The \emph{software architecture} of a computing system is the set of structures needed to reason about the system, which comprises software elements, relationships among them, and properties of both.}
\end{quotation}

Definitions aside, the principle idea behind software architecture is to allow software engineers to treat system design at a high-level of abstraction, representing a system as a composition of interacting components. Properties of those components and their compositions can then be specified in a way that allows designers to analyze systemic quality attributes and tradeoffs, such as performance, reliability, security, availability, maintainability, and so on~\cite{shawAndGarlan}.

Since its emergence there has been substantial development of foundations, tools, and techniques to aid software architects. These include formal and semi-formal architecture description languages (ADLs) \cite{MedvidovicT98}, architecture-based analyses~\cite{Garlan2006}, architecture reconstruction tools~\cite{DiscoTectTSE}, architecture evaluation methods~\cite{Clements01ATAM}, architecture handbooks~\cite{Buschmann96}, architecture style definition and enforcement~\cite{Garlan00AcmeChapter}, and many others.

With respect to the theme of this paper, a number of salient features of software architecture are particularly important:
\begin{itemize}
    \item{\textbf{Component composition:}} Software architecture represents a system as a composition of components, supporting a high-level view of the system and bringing to the forefront issues of assignment of function to components, component compatibility, protocols of interaction between components, and ways to package component compositions for reuse.
    \item{\textbf{Domain-specific computation models:}} Software architecture allows developers to represent a system using compositional models that are not restricted by the implementation platform or programming language, but can be chosen to match the intuition of designers. Specifically, software architecture allows one to define \emph{architectural styles}, where each style denotes a family of systems that shares a common vocabulary of composition, conforms to rules for combining components, and identifies analyses that can be applied to systems in that family~\cite{shawAndGarlan}. Styles may represent generic computational models such as publish-subscribe, pipe-filter, and client-server. Or, they may be specialized for particular domains~\cite{Monroe99Thesis,Monroe96ICSR}.
    \item{\textbf{Analysis:}} Software architecture allows developers to perform analysis of quality attributes at a systems level. This is typically done by exposing key properties of the components and their interactions, and then using those properties in support of calculations to determine expected component compatibility, performance, reliability, security, and so on~\cite{Garlan2006}. This in turn allows developers to make engineering tradeoffs, for example balancing attributes like fidelity, performance, and cost of deployment to match the particular business context. Additionally, in some cases it is possible to build analytic tools that not only detect problems, but also suggest possible solutions~\cite{Spitznagel2003}.
    \item{\textbf{Reuse:}} Software architecture supports several kinds of reuse. First, architectural styles provide a basis for sharing components that fit within that style~\cite{Monroe99Thesis,Monroe96ICSR}. Modern examples of this include platforms like JEE and frameworks like Eclipse. Second, software architectures permit the definition of reusable patterns that can be used to solve specific problems~\cite{Bass07BookSecondEdition,Buschmann96}. Third, most architectural models support hierarchical description, whereby a component can be treated as a primitive building block at one level of composition, but refined to reveal its own sub-architecture.
    \item{\textbf{Execution support:}} For some architectural styles tools can generate implementations. Typically this is done by using a repository of components that conform to the style, and then compiling the system description into executable code~\cite{Garlan2005}. Additionally, software architectures can be used for run-time monitoring and debugging~\cite{Yan2004a}.
\end{itemize}

These properties suggest that if applied appropriately, software architecture principles, tools, and practices could directly address the five challenges outlined in Section~\ref{sec:problem}. Specifically:
\begin{enumerate}
    \item \textbf{Excessive technical detail:} Architectural models provide a  way to develop, analyze, and execute compositional models at a high level of abstraction, suppressing details of implementation.
    \item \textbf{Inappropriate computational models:} Architectural models can define domain-specific compositional styles to match the computational intuition of end users.
    \item \textbf{Inability to analyze compositions:}  Architectural models, suitably represented and formalized, can be analyzed by tools to gain insight into a system's expected quality attributes and to evaluate tradeoffs between alternative designs based on their support for relevant qualities.
    \item \textbf{Lack of support for reuse:} Architectural models support reuse of components, patterns, styles, and encapsulated subsystems.
    \item \textbf{Impoverished support for execution.} Architectures can, in principle, be used as a basis for compilation, deployment, execution, and debugging.
\end{enumerate}

How can these potential benefits be realized? We would argue that the key to doing this is to use an approach in which there is an explicit architectural representation of the compositions created by end users. For a given domain the architectures that could be created would be associated with a domain-specific architectural style corresponding to natural computational models for the domain (such as some variant on workflow, publish-subscribe, or data-centric styles). Further, associated with the style and corresponding infrastructure, there would be a set of architecture services that could support analysis, execution, etc. Finally, all of these features would be made available to users through a graphical front end that supports access to component repositories, architecture construction, system execution, and various additional support services.

\begin{figure}[h!tb]
\begin{center}
\includegraphics[scale=0.8]{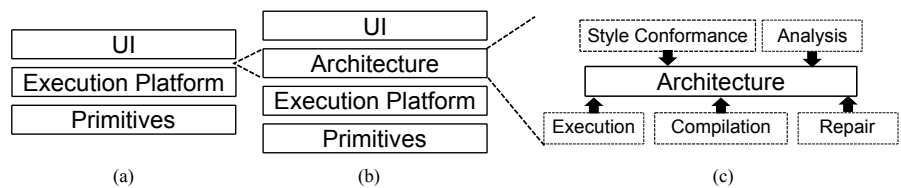}
\caption{End-user Architecting Approach}
\label{fig:Approach}
\end{center}
\end{figure}

This leads to a general framework of system organization in support of end-user architecting, as illustrated in Figure~\ref{fig:Approach}. Part (a) of the figure shows the current state of affairs: users must translate their tasks into the computational model of the execution platform, and become familiar with the low-level details of that platform and the primitive computational elements (applications, services, files, etc.) --- leading to the problems outlined in Section~\ref{sec:problem}.  Part (b) illustrates the new approach. Here, end-user architectures are explicitly represented as architectural models defined in a domain-specific architectural style. These models and the supporting infrastructure can then support a host of auxiliary services, including checking for style conformance, quality attribute analysis, compilation into efficient deployments, execution and debugging mechanisms, and automated repair --- as shown in part (c).

\section{Case Studies}
\label{sec:caseStudies}

To investigate the potential of this approach we instantiated the general framework described above in three domains: dynamic network analysis, brain imaging, and geospatial analysis. For each we describe the nature of the domain and the forms of composition that are required within the community of use. We then consider how we adapted the end-user architecting framework to this domain in terms of (a) architecture representation, (b) architecture style, (c) architectural analysis, (d) execution support, (e) additional services, (f) reuse, and (g) user interface.

\subsection{Dynamic Network Analysis}

Dynamic Network Analysis (DNA) is a domain of computation that focuses on the analysis of network models, which represent entities, relations, and their properties. DNA is increasingly being used in a variety of fields, including anthropology, sociology, business planning, law enforcement, and national security, where networks capture the relationships between people, knowledge, tasks, locations, etc.~\cite{carley2006}.

End users in these fields are typically analysts who extract entities and relations from unstructured text (such as web sites, blogs, twitter feeds, email, etc.) to create  network models, and who then use those models to gain insight into social,  organizational, and cultural phenomena through analysis and simulation.

For example, an analyst interested in understanding disaster relief after the Haiti earthquake in 2010~\cite{Zhao2011}  might build a network from open source news data provided through a source such as LexisNexis~\cite{lexisnexis}.
This unstructured textual data needs to be processed into a usable form, or ``cleaned,'' to filter out headers, remove noise, and normalize concepts. From this processed data a dynamic network can be generated representing associations between people, places, resources, knowledge, tasks, and events. Using network analysis algorithms, insights can then be gained. For example, analysis can determine things like the primary organizations and people involved in the relief effort, how information about food and medical supplies propagated through the network, and how these evolved over time.

Similar kinds of analyses are routinely carried out in law enforcement (where analysts use crime reports and statistics to determine drug-related gang activities), healthcare and disease control (where analysts use medical reports from hospitals and pharmacies to understand disease vectors), and anthropology (where social scientists can understand belief systems and how they relate to demographics).

\begin{figure} [h!tb]
\begin{center}
\includegraphics[scale=.5]{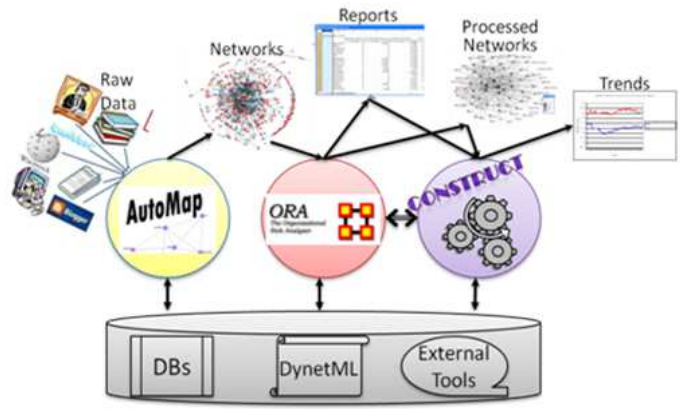}
\caption{Typical tools for socio-cultural analysis.}
\label{toolChain}
\end{center}
\end{figure}

Within this broad domain of dynamic network analysis, analysts typically engage in a process of composing a variety of existing tools to extract networks, analyze them, and display results. Figure~\ref{toolChain} illustrates a typical toolset used for such analyses consisting of the following:  AutoMap for extracting networks from natural language texts, ORA for analyzing and visualizing networks, and Construct for ``what-if'' reasoning about the
networks using simulation~\cite{SchmerlGDBC11}.

Conceptually the computations that analysts create can be viewed as workflows, where each step in the workflow requires the invocation of some data transformation step that consumes the data from previous steps and produces results for the next step. However, traditionally, to achieve this kind of composition analysts would need to understand the idiosyncracies of each of tool, manually invoke them on data stored in various file locations using a variety of file naming schemes and data formats, and preserve the results of the analysis in some location that they would have to keep track of, before invoking another tool to carry out the next step.

More recently coarse-grained tools like AutoMap, ORA, and Construct have been reengineered to expose a set of services that can be composed within a SOA framework. While the use of services reduces the burden of learning to use specific tools, and opens up the possibility of novel compositions, unfortunately the use of SOA requires end users to translate their workflow intuitions into the low-level encodings and scripting required by SOA orchestration languages such as BPEL. Figure~\ref{fig:BPEL} illustrated the resulting complexity of such encodings.

\begin{figure} [h!tb]
\begin{center}
\vspace{-10pt}
\includegraphics[scale=.3]{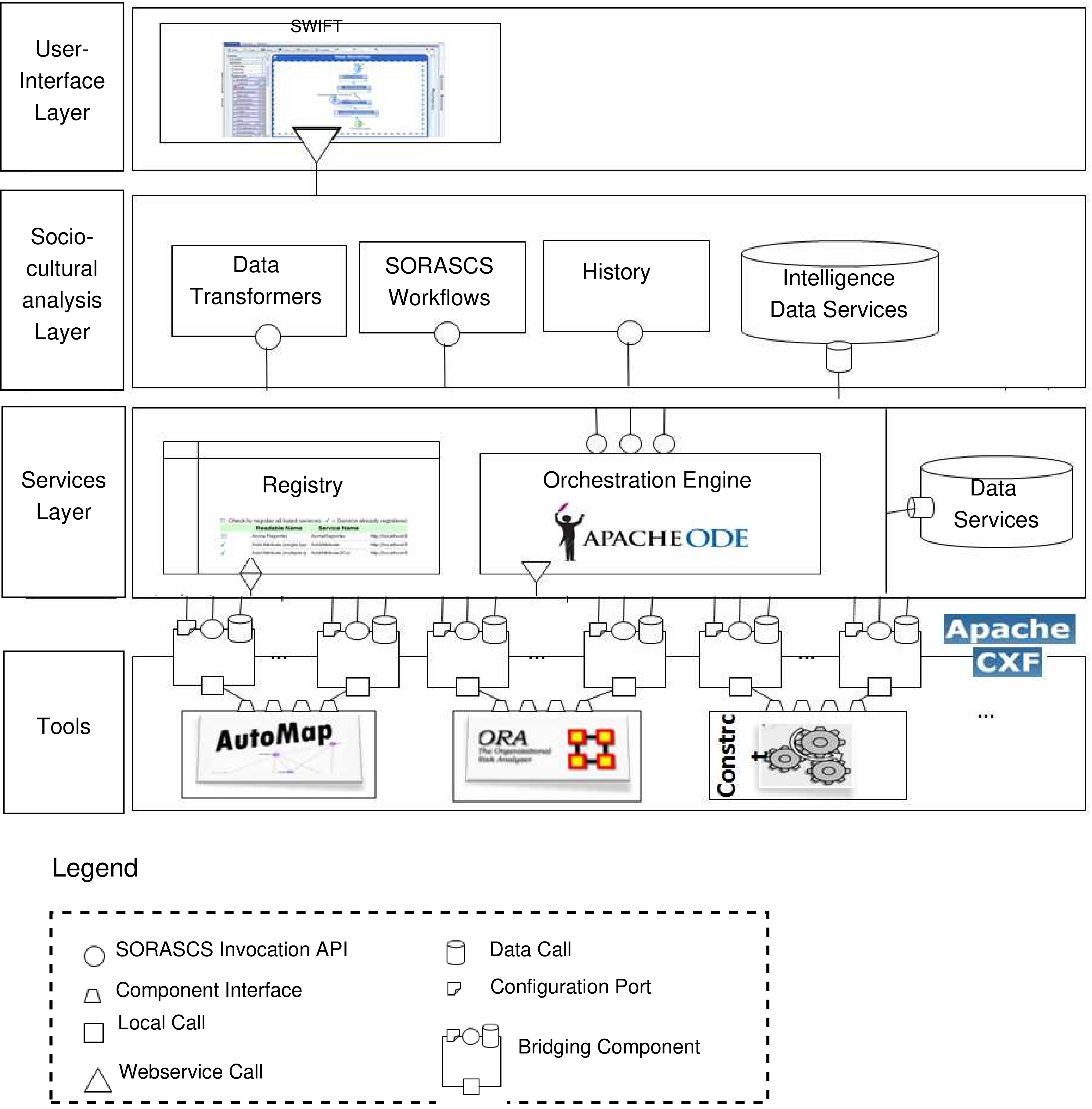}
\vspace{-5pt}
\caption{SORASCS Organization.}
\label{sorascsOrganization}
\end{center}
\vspace{-10pt}
\end{figure}

To apply the proposed end-user architecting approach to this domain, we adapted the end-user architecting framework of Figure~\ref{fig:Approach} by creating an environment, called SORASCS (Service ORiented Architecture for Socio-Cultural Systems), for dynamic network analysis~\cite{Garlan/2009/SOA,SchmerlGDBC11}, and illustrated in Figure~\ref{sorascsOrganization}. Key features of this environment are as follows:
\begin{enumerate}
\renewcommand{\theenumi}{\alph{enumi}}
\item \textbf{Architecture representation:} Architectures are explicitly represented in a system layer, called the socio-cultural analysis layer. This layer stores compositions as workflows. It also provides a repository of data transformers, which act as component building blocks for creation of new workflows.
\item \textbf{Architecture style:} Compositions are defined using a formal workflow architectural style, which specifies the vocabulary of element types and constraints on compositions~\cite{Dwivedi11}. Element types include data transformers, data sources, and data sinks. Constraints of the workflow style prohibit the introduction of cycles, dangling connectors, unattached interfaces, and mismatched communication channels (where the data produced by one component is incompatible with the data consumed by a successor component).
\item \textbf{Analysis:}  The SORASCS workflow style supports a number of analyses including (a) data privacy analysis, which identifies potential privacy issues in the information flows, (b) ordering analysis, which uses machine-learning to evaluate whether the ordering of transformation steps is consistent with previously constructed workflows, and (c) performance analysis, which estimates the amount of time that will be taken to complete an analysis of a specified data set.
\item \textbf{Execution support:} Workflows are compiled into BPEL scripts, which are run within the Services Layer using standard SOA infrastructure. The compilation process attempts to optimize performance by parallelizing workflow execution. Additionally, there is execution support for long-duration transformations and graceful error handling --- typically not provided by baseline SOA infrastructure. Further, it is possible for a user to set breakpoints, execute the workflow one transformation at a time, and preserve intermediate data for later inspection.
\item \textbf{Services:} The SORASCS platform provides services for examining history and for repeating previously executed activities in the history list. The platform also provides data services  for organizing data into projects and categories, and categorizing the data in ways that are informative to analysts. Access control is provided to check that users have appropriate rights to use data sets and transformations.
\item \textbf{Reuse:} Workflows can be encapsulated as parameterized components for later reuse and adaptation. These are stored in a repository of available data transformers, which may be used as primitives, or ``opened'' to reveal their substructure and possibly edited for new usage contexts.
\item \textbf{User Interface:} A web-based graphical interface, called SWiFT~\cite{Garlan/SWiFT/2011}, is provided for workflow construction, analysis, and execution. Further, the interface provides access to the set of available data transformers, organized hierarchically according to community-based ontologies.
\end{enumerate}

\begin{figure} [h!tb]
\begin{center}
\vspace{-10pt}
\includegraphics[scale=.4]{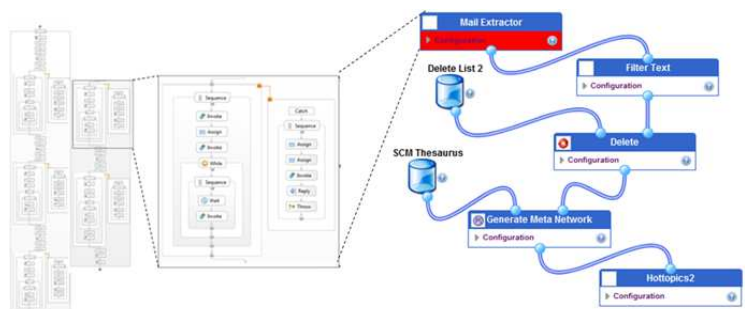}
\vspace{-5pt}
\caption{A DNA Workflow with a Security Flaw.}
\label{fig:simpleWorkflow}
\end{center}
\end{figure}

To illustrate how SORASCS works, Figure~\ref{fig:simpleWorkflow} shows a workflow that analyzes a user's emails to generate a social network of his/her contacts. Table~\ref{tab:dnaOperations} lists the computational elements that are used for this workflow.  The \texttt{Mail Extractor} workflow step acquires security credentials to connect to a remote mail server in order to gain access to the user's emails. The composition then transmits the user's email data  to  \texttt{Filter Text}, followed by \texttt{Delete}, which in combination remove irrelevant words and symbols. This data is then passed to \texttt{Generate Meta-Network}, which generates a social-network of the people and concepts referred to in the email text. \texttt{HotTopics} then creates a report listing important keywords in this social network. The workflow also uses two data sources that provide the inputs to the text processing steps.

When a security analysis is run on this workflow, SORASCS detects a security problem. In this case, data security requirements mandate the use of `token-based authentication' by all services. However the above workflow includes the \texttt{Mail Extractor} service, which uses  `password-based authentication' --- indicating a security violation. The analysis flags this as a problematic workflow by highlighting the inappropriate service in red.

\begin{table}[h!]
\centering
\begin{tabular}{|p{2.5cm}|p{6.5cm}|}
\hline
\textbf{Operation} &	\textbf{Description}	 \\
\hline
\hline
\texttt{Mail Extractor} & 	Extracts email from a server to a text file	 \\
\hline
\texttt{Filter Text}	& Removes undesirable information from text files	 \\
\hline
\texttt{Delete}	&  Removes a set of common keywords using a standard dictionary (such as: a, an, the, etc) from a text file	 \\
\hline
\texttt{Generate Meta-Network}	& Creates a dynamic network based on the information in the text file \\
\hline
\texttt{Hot Topics}	& Creates a report about important keywords in a social network  \\
\hline
\end{tabular}
\vspace{5pt}
\caption{DNA operations used in the workflow of Figure~\ref{fig:simpleWorkflow}.}
\label{tab:dnaOperations}
\end{table}


Once analysis is complete and the errors have been corrected, the user can compile the workflow into the BPEL script illustrated in Figure~\ref{fig:simpleWorkflow}, which can then be executed. Although not illustrated here, as execution proceeds, the user is given feedback through the SORASCS user interface to show which workflow step is currently being executed.

\subsection{Neuroscience}

\begin{figure} [h!tb]
\begin{center}
\vspace{-10pt}
\includegraphics[scale=.3]{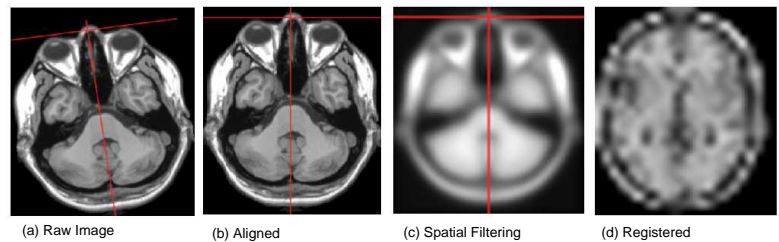}
\vspace{-10pt}
\caption{Brain image data viewed after individual pre-processing steps.}
\label{fig:bImage}
\end{center}
\vspace{-20pt}
\end{figure}

Functional magnetic resonance imaging (fMRI) is a common form of analysis performed by neuroscientists in the brain-imaging domain to understand the behavior of the human brain~\cite{pekar2006}. A typical fMRI analysis consists of sequences of computations over brain image data to support hypotheses or interpretations, such as assessing the evolution of cognitive deficits in neurodegenerative diseases~\cite{eidelberg2009}. Figure~\ref{fig:bImage} illustrates a typical image translation process.

Neuroscientists have at their disposal large repositories of brain imaging data, such as the BIRN Data Repository~\cite{birn} and the Portuguese Brain Imaging Network Project~\cite{bing}. Neuroscientists also have access to a large variety of processing tools, which perform functions such as those listed in Table~\ref{tab:FMRIAnalysis}.

\begin{table}[h!]
\centering
\begin{tabular}{|p{2.5cm}|p{6.5cm}|p{1.5cm}|}
\hline
\textbf{Operation} &	\textbf{Description}	& \textbf{ Tool name} \\
\hline
\hline
\emph{Align}	&	Alignment of an fMRI sequence based on a reference volume (i.e. motion correction, direction correctness) & \textit{fslmaths, fslroi, mcflirt} \\
\hline
\emph{Segmentation}	& Segmentation of a brain mask from the fMRI sequence &	 \textit{bet2, fslmaths, fslstats} \\
\hline
\emph{Spatial Filtering}	& Compute spatial density estimates for neuroscience images, and filter the volumes accordingly	& \textit{fslmaths, susan} \\
\hline
\emph{Temporal Filtering} & Blur the moving parts of images, while leaving the static parts.	& \textit{fslmaths} \\
\hline
\emph{Normalize}	& Translating, rotating, scaling, and may be wrapping the image to match a standard image template &	\textit{flirt} \\
\hline
\emph{Register}	& Align one brain volume to another using linear transformation operations (such as rotation, translations, etc.) or non-linear transformations (such as warping, local distortions, etc.) &	 \textit{flirt, fnirt} \\
\hline
\end{tabular}
\vspace{5pt}
\caption{Some tools for brain-imaging processing.}
\label{tab:FMRIAnalysis}
\vspace{-10pt}
\end{table}

Professional neuroscientists can easily identify the steps required for processing brain imaging data, but because of a proliferation of possible tool implementations for each step and their idiosyncratic parameterization requirements, they find it difficult to choose and assemble tools to implement these steps. Furthermore, while these experts can debug a processing script by examining the outputs, novices are typically unable to do this. As an example of the complexity introduced by tool parameterization, Figure~\ref{fig:neuroscience} illustrates a part of a typical script in which a single logical processing step requires the specification of 9 parameters\footnote{In practice, the number of parameters ranges from 5 to 25.}.


Additional complexity arises because of implicit sequencing constraints. For example, a mandatory step in fMRI analysis is to perform pre-processing operations on brain image data to remove or control some aspects that can affect the overall analysis~\cite{strother2006} (such as aligning one brain volume to another using linear transformations operations like rotation, translation, etc.). While experts may learn these constraints through trial and error, there are no tools to guide less-expert end users.


There are many possible ways to encode image data and analysis results, and neuroscientists must ensure that encodings match between steps. This further complicates composition because neuroscientists must be aware of these formats and carefully select compatible steps or manually locate transducers that can bridge mismatches.

\begin{figure} [h!tb]
\begin{center}
\vspace{-10pt}
\includegraphics[trim=0cm 0cm 1.3cm 0cm, clip=true, totalheight=0.3\textheight]{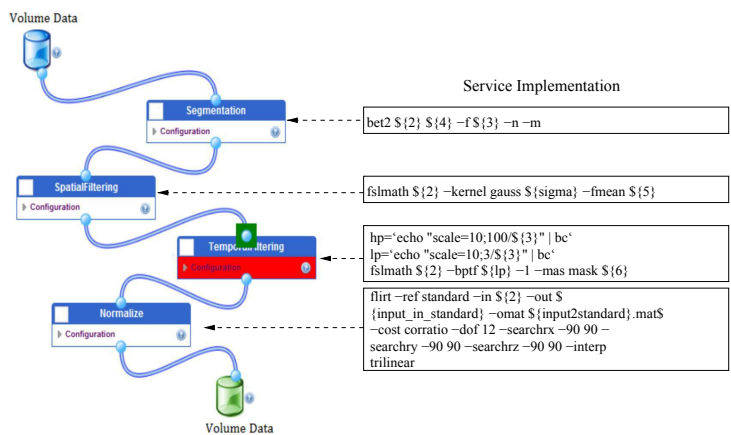}
\vspace{-5pt}
\caption{A problematic neuroscience workflow that misses `alignment' of data before `temporal filtering'.}
\label{preProcessingWF}
\end{center}
\vspace{-10pt}
\end{figure}

To address these problems we adapted the end-user architecting framework  to this domain as follows:
\begin{enumerate}
\renewcommand{\theenumi}{\alph{enumi}}
\item \textbf{Architecture representation:} Similar to dynamic network analysis, architectures are explicitly represented in a system layer that stores compositions as workflows and provides a repository of processing steps and transducers. The main components made available in this prototype were derived from the FSL tool suite (e.g., \textit{bet2, fslmath, flirt})~\cite{fsl}.
\item \textbf{Architecture style:} Compositions are defined using a formal workflow architectural style, which is similar to the one used for dynamic network analysis.\footnote{In fact, using the formal architectural description language of Acme\cite{MonroeKMG97}, we have defined a common root style for both the dynamic network analysis domain and the neuroscience domain~\cite{Dwivedi11}.} The neuroscience style differs in two respects: (a) it defines computational elements specific to the neuroscience domain, and (b) it provides additional properties and domain-specific constraints (such as checking ports for different data encodings and other content of brain-image data) that allow the correct construction of workflows within the neuroscience domain.
\item \textbf{Analysis:} Similar to dynamic network analysis, the properties of the style elements are used for designing various domain-specific analyses for the brain imaging domain. An example is data mismatch analysis to support the detection of data mismatches in the neuroscience compositions and to suggest repairs that can resolve these mismatches based on an end user's quality of service requirements~\cite{Velasco2012}.
\item \textbf{Execution support:} Workflows are compiled into BPEL scripts, which are executed on a service-oriented platform, identical to SORASCS, providing the similar feedback and debugging facilities.
\item \textbf{Services:} Similar to dynamic network analysis, the brain imaging platform provides services to end users tracking the history of operations performed and access to brain imaging data sets.
\item \textbf{Reuse: }Like dynamic network analysis, workflows can be encapsulated as parameterized components for later reuse and adaptation.
\item \textbf{User Interface:} A web-based graphical interface is provided for workflow construction, analysis, and execution.
\end{enumerate}



Figure~\ref{preProcessingWF} illustrates a typical application that analyzes  brain image data using some of the transformation operations listed in Table~\ref{tab:FMRIAnalysis}. To the right of the workflow the figure indicates the invocation and parameter settings that are used to invoke individual tools.

In this example analysis reveals an error in the workflow located in the \texttt{Temporal Filtering} component and its corresponding interface. The error occurs because before doing temporal filtering on  brain-imaging data, it is necessary to align it. Therefore any workflow is required to have the \texttt{Align} component before the \texttt{Temporal} \texttt{ Filtering} component. This is an example of a typical semantic problem that cannot be easily identified from scripts or BPEL-like compositions.

\subsection{Geospatial Analysis}

Geospatial analysis tools allow analysts to explore location-based data using graphical representations such as maps and charts \cite{smith_geospatial_2007}. Examples of such data include data about infrastructure (e.g., an electrical grid), population distribution (e.g., census data), or dynamic network data that has location information associated with it (e.g., crime activities associated with a criminal network derived from police reports). End users in this field typically want to display information on one or more maps, drill down into more detail in certain views, and receive updates when information changes. In contrast to dynamic network analysis and neuroscience analysis, which is largely sequential and transformational, end users doing geospacial analysis typically explore information through a set of concurrent tools that exchange dynamically-changing data to update multiple concurrent views.

The Ozone Widget Framework (OWF)~\cite{potomac_fusion_ozone/synapse_2012} -- or just Ozone -- is a web platform for integrating web-based tools in this domain. Web applications are represented as lightweight visual applications, called \emph{widgets}, and OWF allows end users to open and compose a set of widgets through a web ``dashboard'' in their browser. Users interact with widgets, which communicate among each other using the OWF framework.

\begin{figure} [h!tb]
\begin{center}
\includegraphics[width=\linewidth]{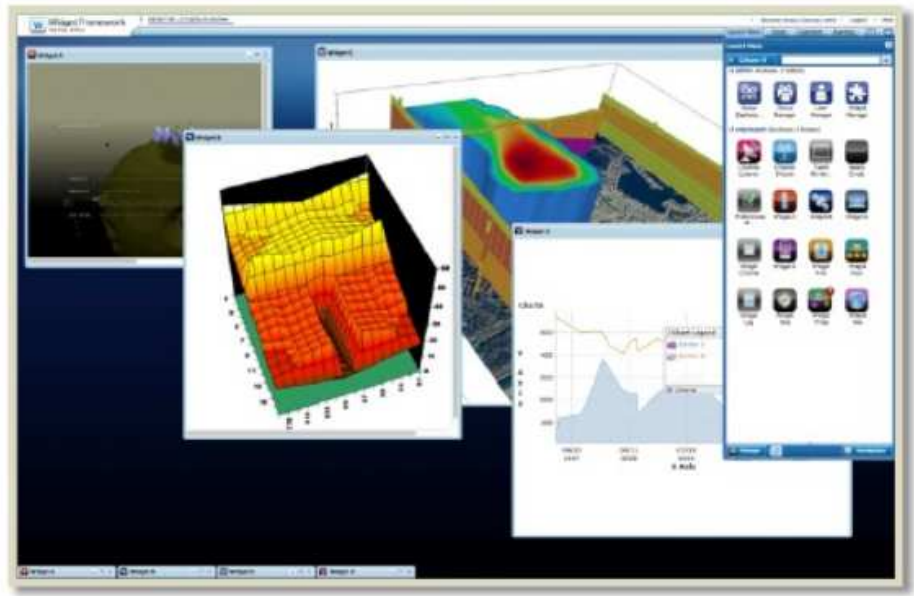}
\caption{An Ozone dashboard example from~\cite{Hellar2012}.}
\label{fig:ozone-map-chart}
\end{center}
\end{figure}

An example of an Ozone dashboard is shown in Figure~\ref{fig:ozone-map-chart}. The right-most window is the launch menu from which end users can add widgets to their dashboard. There are four widgets displayed on the dashboard, displaying information of different types, some in chart form, others (in the background) on maps. These widgets may pass information between each other to ensure that they are focused on the same map region, for example, or to display updated information as it becomes available from a database or data stream. This dashboard and the arrangement of widgets can be shared between developers by exchanging textual configuration files.

%

Ozone widgets interact in a publish-subscribe style~\cite{clements_documenting_2010}: widgets can publish events to channels and subscribe to channels to receive events.\footnote{Events in Ozone are plain-text strings or JSON objects.}
All widgets that have subscribed to a channel receive data published to that channel by any other widget. Widget developers who wish to integrate with other developers must agree on the names of channels to publish to, and the format of the data that is published. To offer additional control over communication, Ozone also allows end users to restrict potential communication between widgets by indicating pairs that are allowed to communicate, thereby implicitly restricting other widgets from participating in those communications.

While end users are free to choose which widgets appear in their dashboard, considerable care must be taken to ensure sensible configurations. In particular, it is important to make sure that widgets both publish and subscribe to the appropriate channels, and that the type of data published is consistent with that expected by subscribers.

Unfortunately, today it is difficult to do this because the interconnection topology is largely implicit. Specifically, to determine the interconnection structure between widgets an end user needs to either examine widget source code, or perform experiments. This problem is compounded by the use of restriction lines, because they can radically change the communication topology indicated in the code by prohibiting interactions that would otherwise be allowed.

The existence of complex interconnection rules and behavior lead naturally to the use of architectural modeling of widget compositions, which could support the end-user architecting process through automated constraint checking. For example, a widget topology can be checked to conform to a privacy constraint that widgets containing private data do not communicate it to third-party untrusted widgets. Another application is widget topology generation: a user would specify what pairs of widgets should and should not interact, and a set of topologies would be generated.

Key features of our end-user architecting approach to this domain are:




\begin{enumerate}
\renewcommand{\theenumi}{\alph{enumi}}
    \item \textbf{Architecture Representation:} Ozone widget configurations are represented as explicit architectural models, that indicate which widgets are involved in a composition and the communication topology.
    \item \textbf{Architectural Style:} Compositions are defined using a variant of a publish-subscribe style that takes into account the idea of restrictions. Element types include Widgets, which have publish and subscribe interfaces, and two types of connectors representing public channels and private (restricted) channels.
    \item \textbf{Analysis:} We are building analyses to provide insight into the widget compositions, such as which widgets are communicating, whether there are data mismatches over publish-subscribe channels, how to restrict communication to minimize event messaging, whether information is lost (e.g., because there is no widget subscribed to information on a particular channel).
    \item \textbf{Reuse:} Dashboard setups (i.e., configurations) can already be shared between analysts as textual configuration files. Embellishing this with architectural representations allows end users to check whether adaptations to existing compositions retain prior communication channels, and whether it is feasible to substitute one widget for another.
    \item \textbf{Services:} Similar to dynamic network analysis,we expect to be able to provide automated data mismatch detection and repair.
    \item \textbf{Execution support:} We are building support for debugging in the form of channel monitoring and execution histories.
    \item\textbf{User interface:} An explicit architectural model enhances the current Ozone user interface by providing information to the end user about which widgets are sharing information with other widgets, which widgets are restricted from communicating, and so on.
\end{enumerate}

\section{Related Work}
\label{sec:relatedWork}

Three primary areas of related research have influenced the formulation and direction of this work: (a) end-user software engineering, (b) software architecture design, and (c) tools and frameworks for end users.

\subsection*{End-user software engineering}

End-user software engineering is a research area at the intersection of computer science and human-computer interaction. It aims to empower users who do not have deep technical expertise to harness the power of computers in support of tasks within their profession~\cite{KoABBBESLLMRRSW11}. Although such users do not have (or want to have) the skills of professional software developers, often they face many of the same software engineering challenges: understanding requirements, carrying out design activities, supporting reuse, quality assurance, etc. In fact, studies have shown that across many domains, such end users spend about 40\% of their time doing programming-related activities~\cite{HowisonH11}, but employ few of the tools and techniques used by modern software engineering. As as result, creating computations often leads to systems that are brittle, contain numerous bugs, have poor performance, cannot be easily reused or shared, and lead to a proliferation of idiosyncratic solutions to similar problems within a domain~\cite{BrandtGLDK09}.


To date, most of the research in end-user software engineering has focused on end-user \emph{programming}, where novel forms of programming languages have been developed for enhanced usability within a domain. These include visual programming languages~\cite{Myers90}, programming-by-demonstration~\cite{cypher-pbe}, direct manipulation programming languages~\cite{Hutchins}, and domain-specific languages~\cite{Fowler11}.

In contrast, this paper focuses on domains in which component composition is the primary form of end-user system construction, an activity that we have termed end-user architecting. For such domains, we have argued, it makes sense to explore ways to adapt the tools and techniques of software \emph{architecture}, rather than software \emph{programming}.



\subsection*{Software architecture}

As we discussed in Section~\ref{sec:endUserArchitecture}, there exists a large body of foundational work on software architecture that has paved the way for architecture to be used as a model to reason about a software system. In this paper we build directly on that heritage. Key influences have been architecture description languages~\cite{MedvidovicT98}, the use of architectural styles~\cite{shawAndGarlan,MonroeKMG97}, and architecture-based analyses~\cite{Garlan2006}.

In this paper we have argued that these techniques have direct relevance and can be effective in solving many of the problems of end-user architecting. However, as we elaborate in Section~\ref{sec:discussion}, there also remain a number of gaps and challenges that require additional research and adaptation of those techniques to the needs of end users.



\subsection*{Tools and frameworks for end-user composition}

The primary motivation for this paper is the fact that a large number of domains require technically-naive users to compose computational elements into novel configurations, such as workflows and scripts for experiments and analyses. Such users often form large communities that share a common set of tasks, vocabulary, and computational needs. These communities include astronomy~\cite{Deelman2005}, bioinformatics~\cite{Letondal2005}, environmental sciences~\cite{VillaAR09}, intelligence analysis~\cite{SchmerlGDBC11}, neuroscience~\cite{neugrid}, and scientific computing~\cite{Segal07}. In such communities simple turnkey or parameterized implementations are inadequate, since it is impossible to anticipate all possible configurations --- hence the need for tools that can help users in creating, executing, and sharing compositions.

As a consequence, a number of powerful composition environments have been created for particular problem domains. Examples include: Loni-pipeline~\cite{Rex&al:03} for brain-imaging compositions; Galaxy~\cite{Giardine:2005} for genomics; and Vistrails~\cite{BavoilCSVCSF05} for data-exploration and visualization for scientific applications.
Other more generic composition environments, such as Taverna~\cite{Oinn2006}, Kepler~\cite{LudascherABHJJLTZ06}, WINGS~\cite{GilRDMK07}, and Ozone~\cite{MooreOzone11}, can be used across several domains, but typically only support a specific computation model --- such as workflow or publish-subscribe.

In contrast to these efforts, this paper attempts to lay the foundation for viewing this class of tools and frameworks as supporting architecture design, and argues that there are considerable benefits in taking this point of view. Among those benefits are the ability to formally define and reason about compositional models as instances of domain-specific architectural styles, create cross-domain analyses, provide systematic support for reuse and adaptation, support powerful auxiliary services (e.g., mismatch repair), and support execution, testing, and debugging.


\section{Discussion}
\label{sec:discussion}

Having described an approach to end-user architecting and illustrated it through three case studies, we now consider some of the aspects of that approach in more detail and outline some of the challenges and open problems.

The centerpiece of an end-user architecting approach is the explicit representation of a composition of computational elements as an architecture, expressed within an appropriate architectural style for the domain at hand. In the case of dynamic network analysis and neuroscience we used variations on a dataflow style. In the case of geospatial analysis we used a publish-subscribe style.

But where does that style come from? In our own experience, we have found that it is often non-trivial to determine this. For example, in the case of dynamic network analysis we found that in some compositions, users wanted to include interactive tools as components in their workflows, in addition to data transformers. This led to a hybrid style that was not purely transformational (as would be the case for a pure dataflow style), but rather permitted a user to interrupt a data transformation workflow, and interactively explore data using applications running on the desktop, before continuing with successive data transformation. Formally, we had to introduce into the style a new type of component --- an interactive tool component --- and create execution infrastructure to permit those components to work smoothly with data transformation executing on a SOA (see~\cite{SchmerlGDBC11} for details).

Similarly, we were initially unsure how to model the communication restrictions present in the Ozone Widget Framework. After exploring a number of options we eventually decided on a variant of a publish-subscribe style that includes two publish-subscribe connector types: public and private pub-sub channels.

The problem of defining an appropriate end-user architecting style is further complicated by the fact that end users may have different compositional needs at different times. For instance, in many analytical domains (including all three domains that we studied), it is the case that in early stages of development end users want to do exploratory investigation using highly interactive, manually-controlled tools. But once it is clear what kinds of computation need to be done, a more streamlined composition can be constructed that provides better performance and is easier for others to use as a packaged computation. This suggests that end users may have several modes of composition, with different architectural modeling needs.

Thankfully, today there are a number of tools that allow one to experiment with different styles. For instance, in our own work we used Acme and its supporting Acme Studio toolset~\cite{Garlan00AcmeChapter}. Acme supports rapid design and experimentation with styles. In particular, styles can be defined using a declarative language, which can then be directly compiled into an environment for constructing systems in that style and for checking conformance with the constraints of the style. Acme Studio also provides an analysis plug-in framework that allows one to rapidly develop analyses appropriate for a given style~\cite{Garlan2006}.

Moreover, Acme has a rich set of base styles (client-server, publish-subscribe, etc.), which can be used as a starting point defining domain-specific styles for end-user architecting communities. For instance, both the dynamic network analysis style and neuroscience style were developed by specializing a common inherited dataflow style. Further, since Acme styles are formally defined they may also be formally analyzed as specifications in their own right to determine, for example, whether a style has the properties that one expects, or to detect inconsistencies when multiple styles are combined~\cite{Kim2010}.

Another technique that helps address this problem is construction of support services that  bridge the gap between different modes of composition. In SORASCS, for example, we provided tools to transition between  interactive exploration and workflow. Specifically, an end user can manually and interactively invoke operations on data sets. SORASCS keeps track of the history of these invocations. Once users are happy with the results, they can use the history to generate a workflow that captures the overall transformation that they want to package as a workflow.

A second concern that must be addressed when pursuing an end-user architecting approach is the issue of managing large component repositories. As we indicated earlier, for many domains there may be hundreds of possible elements that can be combined to produce compositions. In SORASCS, for example, there are over 100 data transformations that are available for dynamic network creation, analysis, visualization, simulation, and report generation. Thus any effective tool for end-user architecting will need to provide scalable ways to search repositories. We have experimented with several schemes for this. For example, we can use community-based ontologies to organize services into categories familiar to end users. We can provide a set of standard filters that can be used to extract components with appropriate properties along several dimensions. We can also use machine learning to recommend possible component selections, based on prior compositions. However, this remains an open problem, as few software architecture tools have addressed the problem of rich component repositories.

A third concern is whether we have raised the level of abstraction sufficiently high. While end-user architecting is a huge improvement over today's programming-based systems, it still requires end users to consider carefully how their computations are composed from the available components. For some users --- particularly novice users, or users who are simply reusing existing compositions --- this may still require too much expertise.

This suggests that in many cases it may make sense to provide another level above that of architecture representation that more directly supports user tasks. For instance, there might be simple domain-specific languages that can be used to define some computation task. Or, there may be simplified interfaces that automatically construct the architectures through various menus or ``wizards''. For example, with SORASCS we demonstrated the ability to do this by connecting it to a front-end tool, called VIBES~\cite{alion_ma&d_operation_vibes_2012}, that provides a specialized interface for constructing belief network analyses. 

More generally, the presence of an intermediate level of architecture simplifies the problem of providing task assistance to end users, since the gap between a task and an architecture that supports it is usually much smaller than the gap between a task and its executable.
However, task-level support for end users seems a particularly rich area for future research, and many questions remain open. For example, is it possible to learn compositions by watching experts solve certain tasks? Can automated synthesis be used to achieve a computational goal based on a high-level description of the inputs and desired outputs?

A fourth concern is the engineering cost for creating end-user architecting environments. Ideally it should be possible to generate large parts of the N-tiered framework that we illustrated in Figure~\ref{fig:Approach}. This remains an open and active area of research.

Finally, as we noted in Section~\ref{sec:problem}, one of the common elements of end-user architecting communities is that they often involve complex ecosystems. In this paper we have primarily addressed only one role within these ecosystems -- the end-user architect. But there are also other roles, such as component developers, data set providers, regulatory bodies, funding agencies, etc.

We have found that when following the end-user architecting approach advocated in this paper, it is also critical that these other roles be considered. For instance, what incentives are there for people to contribute reusable components to an end-user architecting platform? If none are in place, it is unlikely that there will be a sufficiently large base of parts for end users to assemble. Has the platform been constructed in such a way that it can be certified for use in deployment environments where there may be significant privacy or security requirements? If not, the end-user architecting tools may not be usable in the target context. How can an analyst who has created a composition get credit for that design if it is used by others? In many communities people are reluctant to make their tools available or share their analyses unless they receive some professional recognition for doing this.

While the approach we have advocated above does not by itself address the entire ecosystem, it can, however, help address some of the concerns such as those mentioned. For instance, analytical outputs of some computation can be formally linked to the composition that produced  those results, providing a way to acknowledge the developers of the individual components and the composition itself. Additionally, as we have indicated, style-based analyses can guarantee certain properties of a composition --- such as security or privacy. Tools can enforce that such analyses are successfully completed before permitting execution of a composition. Further, the decoupling of the architecture from the execution infrastructure on which it runs allows one to select an execution platform that satisfies regulatory concerns.\footnote{For instance, there are certain pre-approved infrastructures for the US military. By using these, one limits the amount of certification that must be done to the parts that are built on top of it.} That said, the understanding of ecosystems for end-user architecting communities remains a largely unexplored area, and a rich subject for future research.

\section{Conclusion}
\label{sec:conclusion}

We have argued that the computational activities of end users in many domains are analogous to that of software architects, and that rather than forcing end users to become programmers, we should instead provide architecture-based tools and techniques to support their tasks.

To make this concrete, we outlined six  elements of an approach: (a) explicit representation of compositions as architectures, (b) use of domain-specific architectural styles to provide appropriate computational models, (c) the ability to analyze end-user architectures for properties such as performance, reliability, security, etc., (d) support for execution and debugging, (e) support for reuse, and (f) possibly additional services that leverage the architectural representation. We then illustrated how this approach can be used in three end-user architecting domains: dynamic network analysis, neuroscience, and geospatial analysis.

We believe that the recognition of the value of architectural modeling for end users in certain domains is an important first step towards improving the ability for myriad disciplines to leverage the power of computation without requiring its participants to become programmers. However, we also acknowledge that there is much more to be done to make this a reality, and we outlined some of the possible future directions in Section~\ref{sec:discussion}.

\section*{Acknowledgments}
This work was supported in part by the Office of Naval Research grant ONR-N000140811223, and the Center for Computational Analysis of Social and Organizational Systems (CASOS). The views and conclusions contained herein are those of the authors and should not be interpreted as representing the official policies, either expressed or implied, of the Office of Naval Research, or the U.S. government. The authors would like to thank Perla Velasco Elizondo, Jose Maria Fernandes, Diego Estrada Jimenez, Aparup Banerjee, Laura Gledenning, Mai Nakayama, Nina Patel, and Hector Rosas for their contributions to various aspects of this work. 

\bibliographystyle{plain}
\let\oldbibliography\thebibliography
\renewcommand{\thebibliography}[1]{%
  \oldbibliography{#1}%
  \setlength{\itemsep}{.5pt}%
}
\bibliography{bibliography}

\end{document}